\documentclass[sigconf]{acmart}

\usepackage{booktabs} % For formal tables
\usepackage{algorithm}
\usepackage{amsmath}
\usepackage[noend]{algpseudocode}
\usepackage{multirow}
\usepackage{makecell}
\usepackage{scalerel}

\usepackage{tabulary}
\newcolumntype{C}[1]{>{\centering\arraybackslash}p{#1}}
\usepackage{bm}

\usepackage{setspace}

\usepackage{enumitem}
\setlist{nolistsep}

\copyrightyear{2022}
\acmYear{2022}
\setcopyright{acmcopyright}\acmConference[SIGIR '22]{Proceedings of the 45th International ACM SIGIR Conference on Research and Development in Information Retrieval}{July 11--15, 2022}{Madrid, Spain}
\acmBooktitle{Proceedings of the 45th International ACM SIGIR Conference on Research and Development in Information Retrieval (SIGIR '22), July 11--15, 2022, Madrid, Spain}
\acmPrice{15.00}
\acmDOI{10.1145/3477495.3532066}
\acmISBN{978-1-4503-8732-3/22/07}

\begin{document}
\fancyhead{}

\settopmatter{printacmref=true}

\title{Thinking inside The Box: Learning Hypercube Representations for Group Recommendation}

\author{Tong Chen}
\affiliation{%
  \institution{The University of Queensland}
}
\email{tong.chen@uq.edu.au}

\author{Hongzhi Yin}
\authornote{Corresponding author.}
\affiliation{%
  \institution{The University of Queensland}
}
\email{h.yin1@uq.edu.au}

\author{Jing Long}
\affiliation{%
  \institution{The University of Queensland}
}
\email{jing.long@uq.edu.au}

\author{Quoc Viet Hung Nguyen}
\affiliation{%
  \institution{Griffith University}
}
\email{quocviethung.nguyen@griffith.edu.au}

\author{Yang Wang}
\affiliation{
  \city{Hefei University of Technology}
}
\email{yangwang@hfut.edu.cn}

\author{Meng Wang}
\affiliation{
  \city{Hefei University of Technology}
}
\email{eric.mengwang@gmail.com}

\begin{abstract}
As a step beyond traditional personalized recommendation, group recommendation is the task of suggesting items that can satisfy a group of users. In group recommendation, the core is to design preference aggregation functions to obtain a quality summary of all group members' preferences. Such user and group preferences are commonly represented as points in the vector space (i.e., embeddings), where multiple user embeddings are compressed into one to facilitate ranking for group-item pairs. However, the resulted group representations, as points, lack adequate flexibility and capacity to account for the multi-faceted user preferences. Also, the point embedding-based preference aggregation is a less faithful reflection of a group's decision-making process, where all users have to agree on a certain value in each embedding dimension instead of a negotiable interval. In this paper, we propose a novel representation of groups via the notion of hypercubes, which are subspaces containing innumerable points in the vector space. Specifically, we design the hypercube recommender (CubeRec) to adaptively learn group hypercubes from user embeddings with minimal information loss during preference aggregation, and to leverage a revamped distance metric to measure the affinity between group hypercubes and item points. Moreover, to counteract the long-standing issue of data sparsity in group recommendation, we make full use of the geometric expressiveness of hypercubes and innovatively incorporate self-supervision by intersecting two groups. Experiments on four real-world datasets have validated the superiority of CubeRec over state-of-the-art baselines.
\end{abstract}

\begin{CCSXML}
<ccs2012>
   <concept>
       <concept_id>10002951.10003317.10003347.10003350</concept_id>
       <concept_desc>Information systems~Recommender systems</concept_desc>
       <concept_significance>500</concept_significance>
       </concept>
 </ccs2012>
\end{CCSXML}

\ccsdesc[500]{Information systems~Recommender systems}

\keywords{Group Recommendation; Hypercube Representations; Self-supervised Learning}

\maketitle

\section{Introduction}\label{sec:intro}
Group activities are natural habitats for humans, and there is no exception in e-commerce. For example, friends can sign up for the same music festivals on Facebook Events, and online platforms like Meetup allow users to form groups and host activities \cite{yin2020overcoming}. As traditional recommender systems only target at suggesting relevant items to individuals, there is an urgent demand in generating recommendations for a group of users, referred to as \textit{group recommendation}. With the increasingly strong synergy between social and commercial functionalities in contemporary e-commerce platforms (e.g., Yelp, Gowalla, Steam, etc.) \cite{yin2016spatio}, there has been a recent surge in developing solutions to group recommendation~\cite{hu2014deep,cao2018attentive,sankar2020groupim}.

Similar to conventional recommendation, latent factor models exhibit dominant performance in group recommendation. In a general sense, groups and items are mapped to points in the vector space (i.e., embeddings), then the pairwise group-item affinity can be estimated via intuitive distance/similarity metrics. As such, current group recommenders share a core motivation of learning representative group-level embeddings to capture the joint preferences of users. In rare cases where group-item interactions are dense, group representations can be directly learned via standard collaborative filtering models by substituting users with groups \cite{yin2019social}. However, the real group activities are highly sparse \cite{yin2020overcoming,sankar2020groupim}, restricting the informativeness of the learned group representations and hindering recommendations for new groups. Thus, a common practice is to aggregate a group's representation from its members' user embeddings, which are usually learned from the richer user-item interactions. Such aggregation strategies have evolved from heuristics like the least misery and maximum pleasure \cite{amer2009group}, to the more recent attentive sum \cite{cao2018attentive} and social influence-based approach \cite{yin2019social}. %Representative methods include taking the mean of all user embeddings within a group, 

Despite the increasing complexity of these recent preference aggregation approaches, the learned group representations still face a major bottleneck in their expressiveness. This is mainly due to the amount of information lost when compressing multiple user embeddings into one. In the context of embeddings, a group/user is represented as a point in the vector space, and each entry is the exact coordinate in the corresponding dimension. However, while each individual user's preference tend to be stable and can be denoted by a certain value in each embedding dimension, the whole group's overall preference reflects a blend of multiplex personal interests, challenging the capacity of point embeddings to accomodate such diversity. For example, if we associate the embedding dimensions with explicit item features, a group of users may land on different preferred values on the ``price'' dimension. But, it quickly becomes problematic when using a single point for the entire group due to the lack of considerations for such disparate and diverse user preferences, which commonly exist in web applications \cite{nguyen2017argument,li2021preference}. 

In some cases, though higher dimensionality can be assigned to the group embedding to encode more information and alleviate the problem, this does not resolve the issue that a single embedding is unable to precisely reflect the decision-making process within a user group. Essentially, aggregating user preferences mimics the decision-making process \cite{guo2020group} where all group members reach a consensus, reflected by a group-level representation. In reality, the group-level representation should involve synergies of common interests as well as compromises on some personal tastes. For example, for two users that have a large discrepancy on preferred prices, it is more realistic for them to agree upon an approximate price range in between, rather than an exact value. This, unfortunately, cannot be facilitated by existing aggregation schemes where all user preferences are merged into a single point. A point group embedding forces all users to agree upon a fixed decision in each dimension, no matter its value is simply averaged from all users \cite{amer2009group} or biased towards more influential group members \cite{yin2019social,cao2018attentive}. 

To this end, we aim to represent the group-level preferences in a more expressive way, such that the multi-faceted user demands are thoroughly encoded, and the intricate decision-making process among all group members are faithfully preserved. Hence, we advocate addressing the restricted capacity of learned group representations by seeking an alternative to the outgoing point embeddings. Specifically, we introduce the notion of hypercubes as a new way to represent the aggregated group-level preferences in the vector space. In each dimension of the vector space, while a point embedding only has a fixed value, a hypercube covers a range of values that span across a continuous interval \cite{ren2020query2box}. Combining the intervals will shape a ``cube''-like subspace, enabling a hypercube to accommodate innumerable points in the vector space \cite{zhang2021learning}. Compared with the traditional group embeddings, hypercubes enable a higher tolerance of different user preferences. Furthermore, as each dimension now encodes a preference interval, this paradigm is a more realistic reflection of how a group of distinct users give concessions to each other and eventually work out a combined criteria for desired items.   

However, further challenges are in place when unleashing the full potential of hypercube group representations. Firstly, as traditional aggregation schemes are dedicated to learning groups' embedding representations, a new paradigm is desired to effectively summarize all group members' vectorized preferences into a hypercube with minimal information loss in user interests. Secondly, the utilization of group hypercubes voids the applicability of existing point-wise distance measures like Euclidean distance or cosine similarity for group-item pairs, highlighting the need for a revamped distance metric that can precisely quantify and distinguish the affinity between different hypercubes (i.e., groups) and points (i.e., items). Thirdly, the long-standing problem of severe data sparsity with group-level interactions \cite{yin2020overcoming,sankar2020groupim} still persists in our case, which will impede the quality of learned group hypercube representations and amplify the risk of overfitting. 

To address these issues, we present our solution to group recommendation, namely the hyper\underline{cube} \underline{rec}ommender (\textbf{CubeRec}). CubeRec inherits the classic point embeddings for individual users/items whose preferences/attributes are deterministic and can be learned from the relatively rich individual-level interactions. Meanwhile, we innovatively represent each group with the more expressive hypercube, which is first composed by the embeddings of its user members, and then optimized through observed group-level interactions. In CubeRec, we propose two alternative strategies to merge individual user embeddings into group hypercubes, specifically a geometric bounding approach that physically encapsulates all group members' points, and an attentive approach that composes hypercubes by investigating the semantics within user embeddings. To allow for accurate item ranking, we put forward a distance function to measure the distance between each item point and group hypercube. The distance metric accounts for distances from both the inside and outside of a hypercube, thus accomodating different scenarios and maintaining a discriminative characteristic. Furthermore, we make full use of the geometry of group hypercubes and propose a novel self-supervised learning paradigm to tackle the data sparsity in group recommendation. In short, we define an intersection operation between two group representations, and leverage the common users shared between two groups as supervision signals to enrich the information encoded within every hypercube.

In summary, the main contributions of our work are:
\begin{itemize}
	\item We define a new schema based on hypercubes to represent group-level preferences for group recommendation. Hypercubes bypass the limited expressiveness and flexibility of conventional point embeddings, allowing the group representations to encode multi-faceted user preferences while fully imitating a group's decision-making process.
	\item We propose CubeRec, a group recommender that comes with novel designs for learning group-level hypercubes and measuring group-to-item distances. To alleviate data sparsity, CubeRec further incorporates self-supervision by innovatively utilizing the intersection between groups.  
	\item We conduct extensive experiments on four real-world benchmark datasets. Experimental results show that CubeRec yields superior recommendation performance compared with state-of-the-art baselines.
\end{itemize}

\section{Preliminaries}\label{sec:prelim}
In this section, we mathematically define the \textit{hypercube representations} of groups, which are the key building block of CubeRec.

One can think of a hypercube as extending a rectangle into a higher-dimensional space, where each edge of the hypercube corresponds to a real-valued, closed interval on each dimension \cite{ren2020query2box,zhang2021learning}. For group $\mathcal{G}$ which is a subset of users, we define its representation $\mathbb{G}$ via the following center-offset format: 
\begin{equation}\label{eq:group_hypercube}
	\mathbb{G} \equiv \{\textbf{g}\in \mathbb{R}^d: \textbf{c}-\textbf{o} \preceq \textbf{g} \preceq\textbf{c}+\textbf{o}\},
\end{equation}
where $\textbf{c}\in \mathbb{R}^d$ is the geometric center of the hypercube, and $\textbf{o}\in \mathbb{R}^d_{\geq 0}$ is the non-negative offset vector. Both the center and offset are $d$-dimensional vectors learned within the same embedding space. Essentially, in the $z$-th dimension of $\mathbb{R}^d$, $\mathbb{G}$ will have its $z$-th edge spanning across $[c_z - o_z, c_z + o_z]$. For notation convenience, we define the hypercube representation for the $n$-th group $\mathcal{G}_n$ as a tuple $\mathbb{G}_n=(\textbf{c}_n, \textbf{o}_n)$ with its group-specific center $\textbf{c}_n$ and offset $\textbf{o}_n$. We use this tuple as a shorthand for Eq.(\ref{eq:group_hypercube}) in the rest parts of our paper. Compared with point embeddings, relaxing the fixed value in each dimension into a flexible range brings an increased tolerance on the variety among group members. It also reflects a more realistic decision process, as the group does not have to pre-approve a specific value per dimension before ranking candidate items.

\section{The Hypercube Recommender}\label{sec:approach}
To provide an overview, CubeRec first learns user and item embeddings based on user-level interactions with items. Then, we propose our solution to building and learning hypercube representations at the group level. To further compensate for the high sparsity of group-item interactions when learning hypercube representations, we introduce a novel learning objective by leveraging self-supervision signals from overlapping groups. In what follows, we unfold the design details of CubeRec. 

\subsection{Learning User and Item Embeddings}
In CubeRec, we employ LightGCN \cite{he2020lightgcn} to learn user and item embeddings. Compared with earlier collaborative filtering (CF) methods, adopting LightGCN facilitates the modelling of higher-order collaborative signals through message passing, and offers simplicity via the omission of excessive nonlinear components \cite{wang2019neural}. We keep this part brief as it is not our core contribution and can actually be replaced by arbitrary latent factor models.

\textbf{Modelling User-Item Interactions.} Treating each user/item as a distinct node, the user-item interactions can be formulated as a bipartite graph, where we let $\mathcal{N}(u_i)$ and $\mathcal{N}(v_j)$ denote the one-hop neighbor sets of user $u_i$ and item $v_j$, respectively. 
The embedding of an arbitrary user/item is updated by propagating its neighbors' embeddings into $u_i$/$v_j$:
\begin{equation}
	\textbf{u}^{(l)}_i =\!\!\!\! \sum_{j \in \mathcal{G}(u_i)}\eta_{ij}\textbf{v}^{(l-1)}_{j},\,\,\,
	\textbf{v}^{(l)}_j =\!\!\!\! \sum_{i \in \mathcal{G}(v_j)}\eta_{ij}\textbf{u}^{(l-1)}_{i},
\end{equation}
where $l\!\leq\! L$ denotes the propagation layer, $\eta_{ij} \!=\! (\sqrt{|\mathcal{G}(u_i)|\!\cdot\!|\mathcal{G}(v_j)|})^{-1}$ is a graph Laplacian normalization term~\cite{he2020lightgcn,kipf2017semi}, while $\textbf{u}^{(0)}_i$ and $\textbf{v}^{(0)}_j$ are randomly initialized and will be trained via back-propagation. Following \cite{he2020lightgcn}, the final embeddings are obtained via mean pooling across all layers, i.e., $\textbf{u}_i\!=\!\frac{1}{l+1}\sum_{l=0}^{L}\textbf{u}^{(l)}_i\!\in\! \mathbb{R}^{d}$ and $\textbf{v}_j\!=\!\frac{1}{l+1}\sum_{l=0}^{L}\textbf{v}^{(l)}_j \!\in\! \mathbb{R}^{d}$, which are used in the subsequent recommendation phases. 

\textbf{Handling Explicit Social Ties.} Obviously, increasing $L$ can help capture higher-order relationships including implicit social relationships like user-item-user paths, as $L$ essentially controls the number of hops to be reached during the neighborhood aggregation. As group recommendation services are commonly provided on social platforms, the explicitly available social connections between users (e.g., following relationship on social media) carry strong signals for understanding a user's preferences. However, the plain LightGCN does not account for such explicit social ties. In CubeRec, we use $\textbf{R} \in \{0,1\}^{|\mathcal{U}|\times |\mathcal{V}|}$ to denote the interactions between users from set $\mathcal{U}$ and items from set $\mathcal{V}$, and $\textbf{S} \in \{0,1\}^{|\mathcal{U}|\times |\mathcal{U}|}$ to denote the social relation adjacency matrix of all users. Then, we formulate the adjacency matrix of the user-item graph as the following:
\begin{equation}\label{eq:social_GCN}
	\textbf{A} = 
\begin{bmatrix}
	\textbf{S} & \textbf{R}\\
	\textbf{R}^{\top} & \textbf{0}
\end{bmatrix},
\end{equation}
where $\textbf{0}$ is a $|\mathcal{V}| \times |\mathcal{V}|$ matrix consisting of $0$s, and the resulted adjacency matrix $\textbf{A} \in \{0,1\}^{(|\mathcal{U}|+|\mathcal{V}|)\times (|\mathcal{U}|+|\mathcal{V}|)}$ now carries both user-item interactions and user-user social ties. With that, we summarize the socially enhanced LightGCN in the matrix form:
\begin{equation}
	\textbf{E}^{(l)} = (\textbf{D}^{-\frac{1}{2}}\textbf{A}\textbf{D}^{-\frac{1}{2}})\textbf{E}^{(l-1)},
\end{equation}
where $\textbf{D} \in \mathbb{N}^{(|\mathcal{U}|+|\mathcal{V}|)\times (|\mathcal{U}|+|\mathcal{V}|)}$ is a diagonal degree matrix in correspondence to $\mathbf{A}$, while $\textbf{E} \in \mathbb{R}^{(|\mathcal{U}|+|\mathcal{V}|)\times d}$ stacks all user and item embeddings. 

\textbf{User-level Recommendation Loss.} We optimize all user and item embeddings with a distance-based objective on user-level interactions, namely the hinge loss \cite{chen2018pme,hsieh2017collaborative}:
\begin{equation}\label{eq:rec_loss}
	L_{user} = \sum_{(u_i, v_{j^+}, v_{j^-})\in \mathcal{D}_{user}}\!\!\!\!\!\!\max(0,\, \lambda + d_{i j^+} - d_{i j^-}),
\end{equation}
where $\mathcal{D}_{user}$ denotes the training set, $\lambda$ is the safety margin size to be predefined, and $d_{ij} = ||\textbf{u}_i-\textbf{v}_j||_2^2$ is the squared Euclidean distance between the embeddings of $u_i$ and $v_j$. The rationale of the hinge loss is that, in the vector space, each user $u_i$ should aways be closer to a visited item $v_{j^+}$ than an unvisited one $ v_{j^-}$. As such, a training sample is defined as a triple $(u_i, v_{j^+}, v_{j^-})$, where every observed user-item pair $(u_i, v_{j^+})$ is matched with a fixed amount of uniformly sampled negative items $v_{j^-}$ to construct $\mathcal{D}_{user}$. 

\begin{figure*}[t]
\centering
\includegraphics[width=7in]{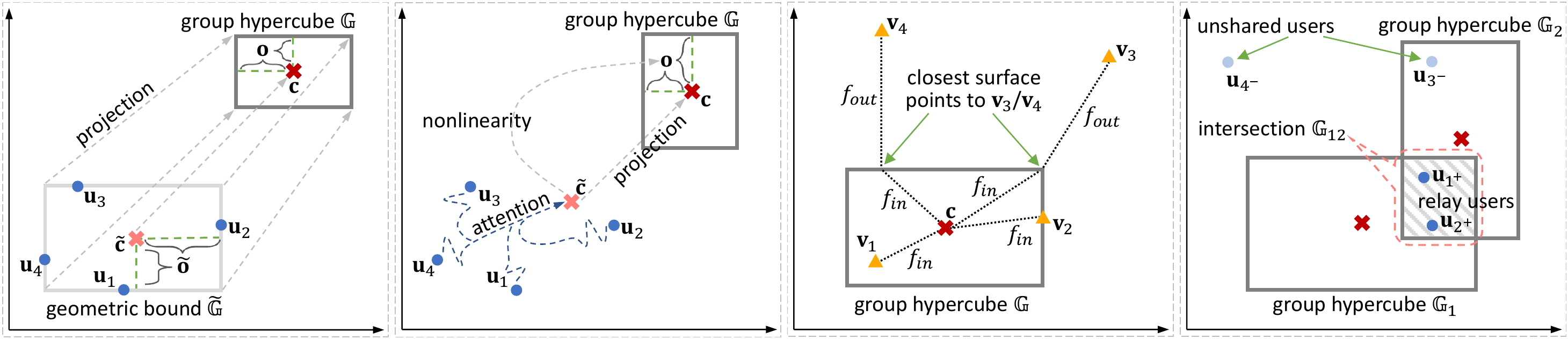}
\small
\begin{tabular}{p{4.1cm}p{4.2cm}p{4.1cm}p{4.1cm}}
	\centering{(a) Geometric fusion and projection for hypercube learning} & \centering{(b) Attentive fusion and projection for hypercube learning} & \centering{(c) Group-to-item distances for different item points} & \centering{(d) Our goal with self-supervised learning via hypercube intersection} \\
	\end{tabular}
\vspace{-0.8cm}
\caption{A schematic view of the key hypercube-based computations in CubeRec. We use $\mathbb{R}^2$ for demonstration purpose. Corresponding details can be found in Section \ref{sec:compose_group} for (a) and (b), Section \ref{sec:rec_loss} for (c), and Section \ref{sec:SSL} for (d).}
\vspace{-0.3cm}
\label{Figure:cube_composition}	
\end{figure*}

\subsection{Composing Group-level Hypercubes}\label{sec:compose_group}
We perform pretraining with $L_{user}$ to acquire initial user and item embeddings for making recommendations at the group level. As a common practice, a group's representation can be learned by merging all the embeddings of its members via means like attention networks \cite{cao2018attentive,huang2020efficient} or graph convolution \cite{wang2020group,guo2021hierarchical}, but they inevitably incur limited expressiveness. As discussed earlier, compressing users' multi-faceted preferences into a single point in the vector space leads to significant information loss, and also misaligns with a group's decision-making process in the real world. 

In this section, we present our solution to learning expressive and flexible group representations with hypercubes. Specifically, we introduce two different strategies for composing group-level hypercubes using individual users' point embeddings, of which both have their unique properties and are further testified in Section \ref{sec:exp}.

\textbf{Geometric Bounding and Projection.}
Given $s$ user embeddings $\textbf{u}_1, \textbf{u}_2, ..., \textbf{u}_s \in \mathbb{R}^d$ in group $\mathcal{G}_n$, the geometric bounding operation is an intuitive way to generate the corresponding group-level hypercube. As Figure \ref{Figure:cube_composition}(a) shows, we first need to find the smallest hypercube covering all $s$ points (i.e., user embeddings) from $\mathcal{G}_n$. Such a hypercube is computed as the following:
\begin{equation}
	\widetilde{\mathbb{G}}_n = (\,\widetilde{\textbf{c}}_n, \widetilde{\textbf{o}}_n) = (\frac{\textbf{u}^{max} + \textbf{u}^{min}}{2}, \frac{|\textbf{u}^{max} - \textbf{u}^{min}|}{2}),
\end{equation}
where $\textbf{u}^{max}$ and $\textbf{u}^{max}$ are boundary vectors computed via:
\begin{equation}
\vspace{-0.1cm}
\begin{split}
	\textbf{u}^{max} &= \max(\textbf{u}_1, \textbf{u}_2, ..., \textbf{u}_s),\\
	\textbf{u}^{min} &= \min(\textbf{u}_1, \textbf{u}_2, ..., \textbf{u}_s),
\end{split}
\end{equation}
where $\max(\cdot)$ and $\min(\cdot)$ both operate element-wise. However, straightforwardly utilizing $\widetilde{\mathbb{G}}_n$ as the final group representation is suboptimal due to its geometric property. That is, the more diverse that the users in $\mathcal{G}_n$ are, the larger area that $\widetilde{\mathbb{G}}_n$ will cover in the $d$-dimensional space. As the hypercube shares the same vector space with items, the recommendation will naturally prefer items that falls in the hypercube because it means those items meet all the preference intervals on all latent dimensions. Consequently, a highly diverse group will receive an excessively large hypercube, covering more false positive items that eventually dilute the recommendation accuracy. To alleviate this issue, we learn projection matrices for both the center and offset that help rescale $\widetilde{\mathbb{G}}_n$ into the final group hypercube $\mathbb{G}_n$:
\vspace{-0.1cm}
\begin{equation}
	\mathbb{G}_n=(\textbf{c}_n, \textbf{o}_n)=(\textbf{W}_c\widetilde{\textbf{c}}_n,\textbf{W}_o\widetilde{\textbf{o}}_n), 
\end{equation}
where $\textbf{W}_c \in \mathbb{R}^{d\times d}$ and $\textbf{W}_o \in \mathbb{R}^{d\times d}_{\geq 0}$ are the corresponding projection weights. Note that $\textbf{W}_o$ is non-negative such that $\textbf{o}_n \geq 0$.

\textbf{Attentive Fusion and Projection.}
Geometric bounding aims to let a hypercube fairly cover all users' preferences within a group. While this is reasonable in many cases, recent studies also point out that each user in the group may affect the final group decision to a different extent \cite{cao2018attentive,yin2019social} due to varied social influence and/or expertise. To make our learned group representations able to account for every group member's importance discrepancy, we further propose an attentive approach \cite{vaswani2017attention,chen2020sequence} for composing group hypercubes: 
\begin{equation}\label{eq:attn_center}
\vspace{-0.1cm}
	\widetilde{\textbf{c}}_n = \textbf{W}_V\textbf{U}_n \cdot \mathrm{softmax} \Big{(}\frac{(\textbf{W}_K\textbf{U}_n)^{\top} \cdot \textbf{q}}{\sqrt{d}} \Big{)},
\end{equation}
where $\textbf{q}\in\mathbb{R}^d$ is the learnable query vector in the above self-attention, $\textbf{W}_K, \textbf{W}_V \in \mathbb{R}^{d\times d}$ are respectively the key and value projection matrices, and $\textbf{U}_n\in \mathbb{R}^{d \times s}$ stacks all $d$-dimensional user embeddings in group $\mathcal{G}_n$. As each $\textbf{u}\in \textbf{U}_n$ reflects a user's exact interest, the aggregation in Eq.(\ref{eq:attn_center}) attentively aggregates all group members' interests, denoted by $\widetilde{\textbf{c}}_n$. Then, we locate the centroid and offset of the composed hypercube $\mathbb{G}_n$ via:
\begin{equation}\label{eq:attn_cube}
	\mathbb{G}_n = (\textbf{c}_n,\textbf{o}_n) = (\textbf{W}_c\widetilde{\textbf{c}}_n,\, \textnormal{ReLU}(\textbf{W}_o\widetilde{\textbf{c}}_n + \textbf{b}_o)\,),
\end{equation}
where we slightly abuse notations $\textbf{W}_c, \textbf{W}_o \in \mathbb{R}^{d\times d}$ to denote the learnable matrices, and $\textbf{b}_o \in \mathbb{R}^d$ is the bias vector. Since the attentively aggregated $\widetilde{\textbf{c}}_n$ summarizes the core interest of the group, we directly use its linear projection as the center $\textbf{c}_n$. Meanwhile, the process of obtaining $\widetilde{\textbf{c}}_n$ sacrifices users' preference diversity, which should be preserved by the offset. Thus, we employ the more expressive nonlinearity to partially infer and recover such missing information from $\widetilde{\textbf{c}}_n$ and obtain the offset $\textbf{o}_n$. The rectified linear unit (ReLU) is used to ensure non-negativity. This process is depicted by Figure \ref{Figure:cube_composition}(b). In short, the attentive approach focuses more on the semantics of user embeddings compared with its geometric counterpart that retains more physical meanings, and is relatively more efficient.

\subsection{Learning Hypercubes for Group Recommendation}\label{sec:rec_loss}
To optimize the composed hypercube representations, we firstly put forward a distance function that quantifies the affinity between each pair of group and item, and then introduce how we define the loss function at the group level for learning CubeRec.

\textbf{Computing Item-to-Group Distances.} Compared with conventional point group embeddings, each edge of a hypercube defines a range on the corresponding dimension in the latent feature space rather than a single value. Intuitively, for any item point $\textbf{v}_j$ that falls in the hypercube $\mathbb{G}_n$, it means each dimension of $\textbf{v}_j$ satisfies the preference range specified by $\mathbb{G}_n$, making item $v_j$ a good fit for group $\mathcal{G}_n$. However, it is not always the case in practice. On one hand, a diverse group tend to have a relatively large coverage with the generated hypercube, containing more irrelevant items. On the other hand, a group with relatively narrow interests can have a rather small hypercube, making it hard to cover any existing item embeddings, especially when $d$ is large. In light of this, we propose an approach that measures the item-to-group distance $d_{nj}$ via: 
\begin{equation}\label{eq:dnj}
	d_{nj} = f_{out}(\mathbb{G}_n,\textbf{v}_j) + \gamma f_{in}(\mathbb{G}_n,\textbf{v}_j),
\end{equation}
where functions $f_{out}(\cdot,\cdot)$ and $f_{in}(\cdot,\cdot)$ measure the outer and inner item-to-group distances, respectively, which are balanced by a fixed coefficient $\gamma \geq 0$. Taking the group hypercube $\mathbb{G}_n$ and item embedding $\textbf{v}_j$ as inputs, these two functions are formulated as follows using squared Euclidean distance:
\begin{equation}\label{eq:fout_fin}
\begin{split}
	f_{out}(\mathbb{G}_n,\textbf{v}_j) &= ||\max(\textbf{v}_j-\textbf{g}^{\triangleright}_n, \textbf{0})+\max(\textbf{g}^{\triangleleft}_n - \textbf{v}_j,\textbf{0})||_2^2,\\
	f_{in}(\mathbb{G}_n,\textbf{v}_j) &= ||\textbf{c}_n - \min(\textbf{g}^{\triangleright}_n, \max(\textbf{g}^{\triangleleft}_n, \textbf{v}_j))||_2^2,\\	
\end{split}
\end{equation}
where two points $\textbf{g}^{\triangleleft}_n,\textbf{g}^{\triangleright}_n \in \mathbb{R}^d$ are geometrically the lower-left and upper-right corners of hypercube $\mathbb{G}_n$, respetively:
\begin{equation}\label{eq:left_right_corner}
	\textbf{g}^{\triangleleft}_n = \textbf{c}_n - \textbf{o}_n, \,\,\,\,\, \textbf{g}^{\triangleright}_n = \textbf{c}_n + \textbf{o}_n.
\end{equation}
As explained in Figure \ref{Figure:cube_composition}(c), assuming $\textbf{v}_j$ locates outside $\mathbb{G}_n$, what Eq.(\ref{eq:fout_fin}) does is to: (1) locate the anchor point on $\mathbb{G}_n$'s surface that is the closest to $\textbf{v}_j$; (2) use $f_{out}(\cdot,\cdot)$ to measure the outer distance from $\textbf{v}_j$ to this anchor point; and (3) use $f_{in}(\cdot,\cdot)$ to measure the inner distance from the centroid $\textbf{c}_n$ to this anchor point. The outer and inner distances are then combined into $d_{nj}$ via Eq.(\ref{eq:dnj}). When $\textbf{v}_j$ is inside or on the surface of $\mathbb{G}_n$, $f_{out}(\mathbb{G}_n,\textbf{v}_j)=0$, and $f_{in}(\mathbb{G}_n,\textbf{v}_j)$ will reduce to the distance between item and centroid because $\min(\textbf{g}^{\triangleright}_n, \max(\textbf{g}^{\triangleleft}_n, \textbf{v}_j)) = \textbf{v}_j$. This is also shown in Figure \ref{Figure:cube_composition}(c).

\textbf{Group-level Recommendation Loss.}
Following the definition of item-to-group distance in Section \ref{sec:compose_group}, the higher the affinity between group $\mathcal{G}_n$ and item $v_j$, the smaller the distance $d_{nj}$ will be. Analogously to Eq.(\ref{eq:rec_loss}), we adopt the following hinge loss:
\begin{equation}\label{eq:group_loss}
	L_{group} = \sum_{(\mathcal{G}_n, v_{j^+}, v_{j^-})\in \mathcal{D}_{group}}\!\!\!\!\!\!\max(0,\, \lambda' + d_{n j^+} - d_{n j^-}),
\end{equation}
where $\lambda'$ denotes the safety margin, $(u_i, v_{j^+}, v_{j^-}) \in \mathcal{D}_{group}$ are training instances constructed by negative sampling. Essentially, for cube $\mathbb{G}_n$, $L_{group}$ aims to pull a positive item's embedding $\textbf{v}_{j^+}$ into $\mathbb{G}_n$ and closer to its center, while pushing the negative item's embeddings $\textbf{v}_{j^-}$ away from it.

\subsection{Self-supervision via Hypercube Intersections}\label{sec:SSL} 
Unfortunately, as pointed out by prior studies \cite{yin2019social,sankar2020groupim}, a long-lasting obstacle for learning high-quality group representations is that, the learning of group representations is purely dependant on the interacted items, which are highly sparse at the group level (see statistics in Section \ref{sec:dataset}). As such, we propose to take advantage of the rich self-supervision signals from user-group associations for learning representative group hypercubes. Specifically, for group $\mathcal{G}_n$ we firstly draw a different group $\mathcal{G}_{n'}$, where $\mathcal{G}_{n'}$ shares at least one user in common with $\mathcal{G}_n$ (details on how we deal with disjoint groups will follow). Users shared between $\mathcal{G}_{n}$ and $\mathcal{G}_{n'}$ are termed \textit{relay users}. Then intuitively, the intersection of two group hypercubes denoted by $int(\mathbb{G}_{n},\mathbb{G}_{n'})$, which is also a hypercube, is supposed to reflect the properties of all the relay users. 

With Figure \ref{Figure:cube_composition}(d), we illustrate our goal of self-supervised learning in CubeRec. In a geometric sense, two hypercubes' intersection is ideally the subspace that hosts the point embeddings of relay users. On this basis, we formulate the self-supervision loss below:
\begin{equation}\label{eq:ssl}
	L_{self} = \sum_{\forall \mathcal{G}_n}\, \sum_{\substack{\forall u_{i^+} \in (\mathcal{G}_n \cap \mathcal{G}_{n'})\\ u_{i^-}\sim p(u_{i^-}|\mathcal{G}_n, \mathcal{G}_{n'})}}\!\!\!\!\!\!\max(0,\, \lambda '' + d_{nn'i^+} - d_{nn'i^-}),
\end{equation}
where $d_{nn'i}$ is a shorthand for the distance between the intersection and user $u_i$, i.e., $f_{out}(int(\mathbb{G}_{n},\mathbb{G}_{n'}),\textbf{u}_i) + \gamma f_{in}(int(\mathbb{G}_{n},\mathbb{G}_{n'}),\textbf{u}_i)$. For each positive user $u_{i^+}$ that falls in the group intersection $\mathcal{G}_n \cap \mathcal{G}_{n'}$, we sample a negative user $u_{i^-}$ from $\mathcal{U}\backslash(\mathcal{G}_n \!\cap\! \mathcal{G}_{n'})$ with uniform distribution $p(u_{i^-}|\mathcal{G}_n, \mathcal{G}_{n'}) = |\mathcal{U}\backslash(\mathcal{G}_n\!\cap\!\mathcal{G}_{n'})|^{-1}$. By doing so, the representations of groups $\mathcal{G}_n$ and $\mathcal{G}_{n'}$ will be regularized by the relay users, so as to capture the common interests between them. Correspondingly, the learned group representations will be substantially more informative than those learned purely with sparse item-level interactions.

\textbf{Hypercube Intersection Operation.} We hereby define how the hypercube intersection $int(\mathbb{G}_n, \mathbb{G}_{n'})$ is calculated. For convenience, we let $int(\mathbb{G}_n, \mathbb{G}_{n'})=\mathbb{G}_{nn'} = (\textbf{c}_{nn'}, \textbf{o}_{nn'})$. Because straightforwardly using the geometric intersection is only applicable when two hypercubes have actual overlaps in each dimension of $\mathbb{R}^d$, we resort to a relaxed intersection computation via a neural approach. Firstly, center $\textbf{c}_{nn'}$ is computed by performing element-wise attention over the two group centers:
\begin{equation}\label{eq:intersec_center}
\vspace{-0.1cm}
\begin{split}
\textbf{c}_{nn'} &= \textbf{a}_n\odot\textbf{c}_n + \textbf{a}_{n'}\odot\textbf{c}_{n'},\\
\textbf{a}_* &= \frac{\exp(\phi(\textbf{c}_*))}{\exp(\phi(\textbf{c}_n)) + \exp(\phi(\textbf{c}_{n'}))} \,\,\, \textnormal{for} \, *\in \{n, n'\},
\end{split}
\end{equation}
where $\odot$ denotes element-wise multiplication, $\phi(\cdot):\mathbb{R}^d\mapsto \mathbb{R}^d$ is a multilayer perceptron (MLP) with $d$ neurons throughout all layers. Meanwhile, the offset $\textbf{o}_{nn'}$ is determined via:
\begin{equation}\label{eq:intersec_off}
\textbf{o}_{nn'} = \min(\textbf{o}_n, \textbf{o}_{n'})\odot \textnormal{sigmoid}(\psi(\textbf{o}_n + \textbf{o}_{n'})),
\end{equation}
where $\psi(\cdot):\mathbb{R}^d\mapsto \mathbb{R}^d$ is another MLP with a distinct set of parameters. Essentially, Eq.(\ref{eq:intersec_center}) places the intersection center $\textbf{c}_{nn'}$ somewhere between $\textbf{c}_n$ and $\textbf{c}_{n'}$. Then, with the scaling effect of sigmoid function, Eq.(\ref{eq:intersec_off}) shrinks the offsets of hypercubes $\mathbb{G}_{n}$ and $\mathbb{G}_{n'}$ to obtain the new offset of the intersection $\textbf{o}_{nn'}$.

\textbf{Handling Isolated Groups.} Despite the heavily overlapping nature of user groups on social e-commerce platforms, we must take into account the situation where a group $\mathcal{G}_n$ might be disjoint with all other groups in the dataset. On this occasion, we propose the two following strategies to generate a dummy user group $\widetilde{\mathcal{G}}_{n'}$ having overlapping users with $\mathcal{G}_n$:
\begin{itemize}
	\item[I.] Proportional Swap (PS): With a specified proportion $\rho$, we swap $\rho |\mathcal{G}_n|$ users in $\mathcal{G}_n$ with users uniformly sampled from $\mathcal{U}\setminus\mathcal{G}_n$ with replacement. We denote this as $\widetilde{\mathcal{G}}_{n'} = \textnormal{PS}(\mathcal{G}_n)$.
	\item[II.] Proportional Imputation (PI): With a specified proportion $\rho$, we inject $\rho |\mathcal{G}_n|$ uniformly sampled users from $\mathcal{U}\setminus\mathcal{G}_n$ (with replacement) into $\mathcal{G}_n$. We denote this as $\widetilde{\mathcal{G}}_{n'} = \textnormal{PI}(\mathcal{G}_n)$.
\end{itemize}
The generated $\widetilde{\mathcal{G}}_{n'}$ then serves as $\mathcal{G}_{n'}$ to facilitate self-supervision via Eq.(\ref{eq:ssl}). We adopt a predefined value $\rho=0.5$ in our approach, where we use PS and PI alternately during training. 
 
\begin{algorithm}[!t]\label{alg:train}
\begin{spacing}{0.9}
\small
\caption{Training Procedure of CubeRec}\label{Algorithm:random}
\begin{algorithmic}[1]
\State \textbf{Input:} $\mathcal{U}$, $\mathcal{G}$, $\mathcal{V}$
\State \textbf{Output:} All model parameters collectively referred to as $\Theta$
\State Randomly initialize $\Theta$;
\Repeat
\State Draw a mini-batch from $\mathcal{D}_{user}$;
\State Take a gradient step to update $\textbf{u}$ and $\textbf{v}$ w.r.t. $L_{user}$;
\Until{convergence}\Comment{Pretraining user and item embeddings}
\Repeat
\State Draw a mini-batch from $\mathcal{D}_{group}$ and compute $L_{group}$;
\For{each $\mathcal{G}_n$ in mini-batch}
\If{$\exists \mathcal{G}_{n'}$ s.t. $\mathcal{G}_n \cap \mathcal{G}_{n'} \neq \varnothing$}
\State Compute $L_{self}$ with a sampled $\mathcal{G}_{n'}$;
\Else
\State Generate $\widetilde{\mathcal{G}}_{n'}$ via either $\textnormal{PS}(\mathcal{G}_{n})$ or $\textnormal{PI}(\mathcal{G}_{n})$ by coin-flipping;
\State Compute $L_{self}$ with $\widetilde{\mathcal{G}}_{n'}$;
\EndIf
\EndFor
\State Take a gradient step to update $\Theta$ w.r.t. $L_{group} + \mu L_{self}$
;
\Until convergence
\end{algorithmic}
\end{spacing}
\end{algorithm}

\subsection{Optimizing CubeRec}\label{sec:opt}
As all components of CubeRec are end-to-end differentiable, we learn the model parameters on multiple objectives with the mini-batch stochastic gradient descent algorithm Adam \cite{kingma2014adam}. As depicted by Algorithm 1, we adopt a two-stage training strategy for CubeRec. To be specific, we first obtain pretrained user and item embeddings by optimizing the user-level loss $L_{user}$. Then, we fine-tune the pretrained embeddings and learn rest of the parameters in CubeRec by optimizing towards the combined loss $L_{group} + \mu L_{self}$ ($\mu \geq 0$). In both training stages, we update corresponding model parameters in each iteration and repeat the entire training process until the loss converges or is sufficiently small. 

We tune the hyperparameters using grid search. Specifically, the latent dimension $d$ is searched in $\{16,32,64,128,256\}$, and the distance coefficient $\gamma$ and self-supervision weight $\mu$ are searched in $\{0.1, 0.3, 0.5, 0.7, 0.9\}$. For optimizing the three losses $L_{user}$, $L_{group}$, $L_{self}$, we draw 5 negative samples for each positive ground truth cases to construct corresponding training sets. The safety margins on those losses are set to $\lambda=\lambda'=\lambda''=0.5$ following common practices for hinge loss \cite{zhang2021learning,chen2018pme}. We set the the learning rate to $1\times 10^{-3}$ and batch size to $256$ according to device capacity. To prevent overfitting, we adopt a dropout \cite{srivastava2014dropout} ratio of $0.2$ on all deep layers of CubeRec during training.

%\vspace{-0.1cm}
\section{Experiments}\label{sec:exp}

\begin{table}[b]
\vspace{0.4cm}
\caption{Statistics of experimental datasets.}
\vspace{-0.5cm}
\renewcommand{\arraystretch}{0.9}
\setlength\tabcolsep{2.8pt}
\center
  \begin{tabular}{l r r r r}
    \toprule
     & Meetup & Yelp & Gowalla & Douban\\
    \hline
    \#users & 24,631 & 34,504 & 60,805 & 70,743\\
	\#groups & 13,552 & 24,103 & 78,453 & 109,538\\
	\#items & 19,031 & 22,611 & 8,984 & 60,028\\
	\#user-item interactions & 126,813 & 482,273 & 1,625,817 & 3,422,266\\
	\#group-item interactions & 19,031 & 26,883 & 208,336 & 164,153\\
	average \#items per user & 5.15 & 13.98 & 26.74 & 48.38\\
	average \#items per group & 1.40 & 1.12 & 2.66 & 1.50\\
	average \#groups per user & 4.83 & 3.11 & 3.23 & 7.53\\
	average group size & 8.79 & 4.45 & 2.31 & 4.86\\
    \bottomrule
\end{tabular}
\label{table:Dataset}
\end{table}

\begin{table*}[t!]
%\vspace{-0.4cm}
\caption{Recommendation results. Numbers in bold face are the best results for corresponding metrics.}
\vspace{-0.4cm}
\centering
\setlength\tabcolsep{0.6pt}
  \begin{tabular}{c|cc|cc|cc|cc|cc|cc|cc|cc}
    \hline
     \multirow{3}{*}{Method} & \multicolumn{4}{c|}{Yelp} & \multicolumn{4}{c|}{Douban} & \multicolumn{4}{c}{Gowalla} & \multicolumn{4}{c}{Meetup}\\
     \cline{2-17}
    & \multicolumn{2}{c|}{Recall$@K$} & \multicolumn{2}{c|}{NDCG$@K$} & \multicolumn{2}{c|}{Recall$@K$} & \multicolumn{2}{c|}{NDCG$@K$} & \multicolumn{2}{c|}{Recall$@K$} & \multicolumn{2}{c|}{NDCG$@K$}& \multicolumn{2}{c|}{Recall$@K$} & \multicolumn{2}{c}{NDCG$@K$}\\
    \cline{2-17}
    & $K=10${\,} & $K=20$ & $K=10${\,}& $K=20$ & $K=10${\,} & $K=20$ & $K=10${\,} & $K=20$ & $K=10${\,} & $K=20$ & $K=10${\,} & $K=20$ & $K=10${\,} & $K=20$ & $K=10${\,} & $K=20$\\
    \hline
    NCF-AVG \cite{he2017neuralcol} & 0.0042 & 0.0090 & 0.0018 & 0.0030 & 0.0023 & 0.0042 & 0.0010 & 0.0015 & 0.0146 & 0.0222 & 0.0088 & 0.0108 & 0.0087 & 0.0092 & 0.0074 & 0.0075\\
    FM-AVG \cite{vinh2019interact} & 0.0161 & 0.0249 & 0.0075 & 0.0097 & 0.0032 & 0.0065 & 0.0014 & 0.0023 & 0.0127 & 0.0212 & 0.0061 & 0.0082 & 0.0128 & 0.0232 & 0.0063 & 0.0089\\
    AGREE \cite{cao2018attentive} & 0.0168 & 0.0286 & 0.0073 & 0.0103 & 0.0031 & 0.0047 & 0.0013 & 0.0017 & 0.0125 & 0.0228 & 0.0063 & 0.0088 & 0.0081 & 0.0117 & 0.0033 & 0.0041\\
    SIGR \cite{yin2019social} & 0.0194 & 0.0311 & 0.0082 & 0.0114 & 0.0035 & 0.0056 & 0.0016 & 0.0020 & 0.0143 & 0.0266 & 0.0070 & 0.0099 & 0.0090 & 0.0139 & 0.0038 & 0.0049\\
    CAGR \cite{yin2020overcoming} & 0.0215 & 0.0345 & 0.0106 & 0.0138 & 0.0026 & 0.0072 & 0.0010 & 0.0021 & 0.0169 & 0.0320 & 0.0090 & 0.0129 & 0.0028 & 0.0029 & 0.0012 & 0.0013\\
    GroupIM \cite{sankar2020groupim} & 0.0424 & 0.0535 & 0.0207 & 0.0235 & 0.0212 & 0.0316 & 0.0098 & 0.0124 & 0.0281 & 0.0384 & 0.0113 & 0.0136 & 0.0407 & 0.0525 & 0.0237 & 0.0266\\
     \hline
    \textbf{CubeRec-G}& \textbf{0.0455} & \textbf{0.0592} & \textbf{0.0241} & \textbf{0.0298} & \textbf{0.0243} & \textbf{0.0354} & \textbf{0.0114} & \textbf{0.0167} & \textbf{0.0361} & \textbf{0.0575} & \textbf{0.0171} & \textbf{0.0194} & \textbf{0.0437} & \textbf{0.0616} & \textbf{0.0261} & \textbf{0.0306}\\
    \textbf{CubeRec-A}& 0.0417 & 0.0520 & 0.0206 & 0.0224 & 0.0211 & 0.0309 & 0.0098 & 0.0158 & 0.0343 & 0.0495 & 0.0135 & 0.0164 & 0.0418 & 0.0525 & 0.0202 & 0.0211\\
    \hline   
    \end{tabular}
\label{table:recommendation}
\vspace{-0.2cm}
\end{table*}

In this section, we conduct experiments to verify the effectiveness of CubeRec in group recommendation. Specifically, we aim to answer the following research questions (RQs):
\begin{itemize}
	\item[\textbf{RQ1}:] Is CubeRec the new state-of-the-art? 
	\item[\textbf{RQ2}:] Are the major components proposed in CubeRec effective?
	\item[\textbf{RQ3}:] How does CubeRec perform w.r.t. different group sizes?
	\item[\textbf{RQ4}:] What is the impact of CubeRec's key hyperparameters?
\end{itemize}

%\vspace{-0.2cm}
\subsection{Datasets}\label{sec:dataset}
We adopt four real-world datasets collected from different event-based social networks, namely \textbf{Meetup}, \textbf{Yelp}, \textbf{Gowalla}, and \textbf{Douban}. Both Meetup and Douban contain group events held in different venues respectively in New York and Beijing, where event venues are items to be recommended. Meanwhile, Yelp and Gowalla are typical check-in datasets on different restaurants (i.e., items in our case) based in the US. Since Yelp and Gowalla do not originally contain group information, we follow the widely adopted procedure \cite{sankar2020groupim,yin2019social} to construct group interactions by finding overlaps on both check-in times and social relations. That is, we assume if a set of users who are connected on the social network visit the same venue at the same time, then they are regarded as members of a group, and the corresponding activities are group activities \cite{yin2019social}. The use of explicit social connections and spatiotemporal tags ensures the quality of discovered user groups in Yelp and Gowalla. We provide the key statistics of the four datasets in Table \ref{table:Dataset}. Each dataset is split with a ratio of $8$:$1$:$1$ for training, validation, and test, respectively.

\subsection{Baselines and Evaluation Protocols}\label{sec:baselines_and_eval}
 We compare CubeRec\footnote{We release our implementation at: https://github.com/jinglong0407/CubeRec.git} with the following state-of-the-art baselines:
\begin{itemize}
	\item \textbf{MF-AVG} \cite{vinh2019interact}: This baseline takes the average of all group-wise user representations learned via user-item matrix factorization (MF) \cite{koren2009matrix} to compose group embeddings.
	\item \textbf{NCF-AVG} \cite{he2017neuralcol}: It also represents groups with average pooling on user embeddings, where user embeddings are learned via the neural collaborative filtering instead of MF.
	\item \textbf{AGREE} \cite{cao2018attentive}: This approach utilizes attention networks, thus accounting for different importance of group members when learning group embeddings.
	\item \textbf{SIGR} \cite{yin2019social}: It exploits and integrates users' global and local social influence to improve the group recommendation.
	\item \textbf{CAGR} \cite{yin2020overcoming}: This method firstly learns centrality-aware user representations, and then learns a group recommender via two-stage optimization.
	\item \textbf{GroupIM} \cite{sankar2020groupim}: This state-of-the-art group recommender adopts mutual information maximization between users and groups to overcome the sparse group-level interactions.  
\end{itemize}

It is worth noting that in CubeRec, we implement both the \textit{geometric} and \textit{attentive} approaches described in Section \ref{sec:compose_group} for composing group hypercubes, which are respectively marked as \textbf{CubeRec-G} and \textbf{CubeRec-A}. We set $L=3$ for learning user and item embeddings with LightGCN, and implement the two MLPs for hypercube intersection with a 3-layer setting and ReLU activation. Based on the hyperparameter search described in Section \ref{sec:opt}, we fix $d=64$, $\gamma=0.3$, and $\mu=0.7$ across all datasets. The impact of these hyperparameters to the recommendation performance will be further discussed in Section \ref{sec:hyperparams}. 

We leverage two metrics, namely recall at rank $K$ (Recall$@K$) \cite{chen2019air,chen2021learning} and normalized discounted cumulative gain at rank $K$ (NDCG$@K$) \cite{chen2020try,zhang2021graph} that are widely adopted in recommendation research. We adopt $K=10,20$ where all items unvisited by each group are taken as negative samples for evaluation. In short, Recall$@K$ measures the ratio of the ground truth items that are present on the top-$K$ list, and NDCG$@K$ evaluates whether the model can rank the ground truth items as highly as possible.

\begin{figure}[t!]
\centering
\begin{tabular}{ccc}
	\vspace{-0.2cm}\hspace{-0.2cm}\includegraphics[width=0.7in]{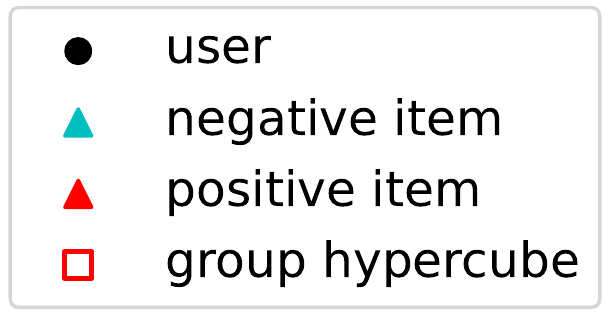}&\hspace{-0.3cm}\includegraphics[width=1.3in]{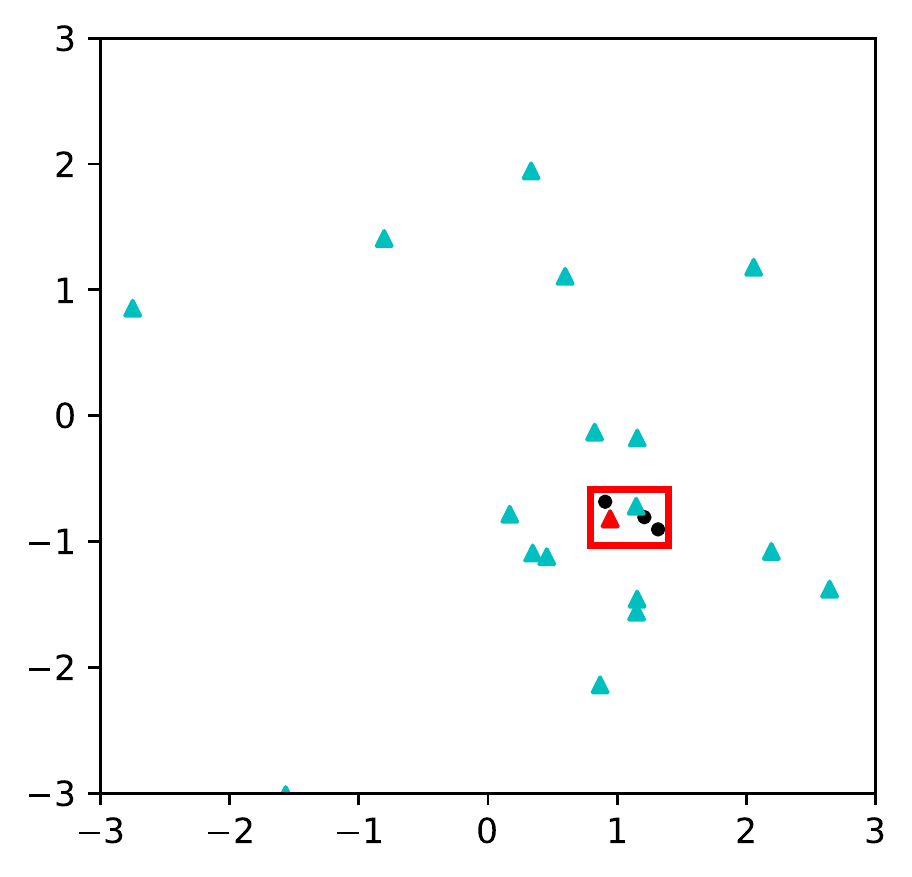}
	&\hspace{-0.15cm}\includegraphics[width=1.3in]{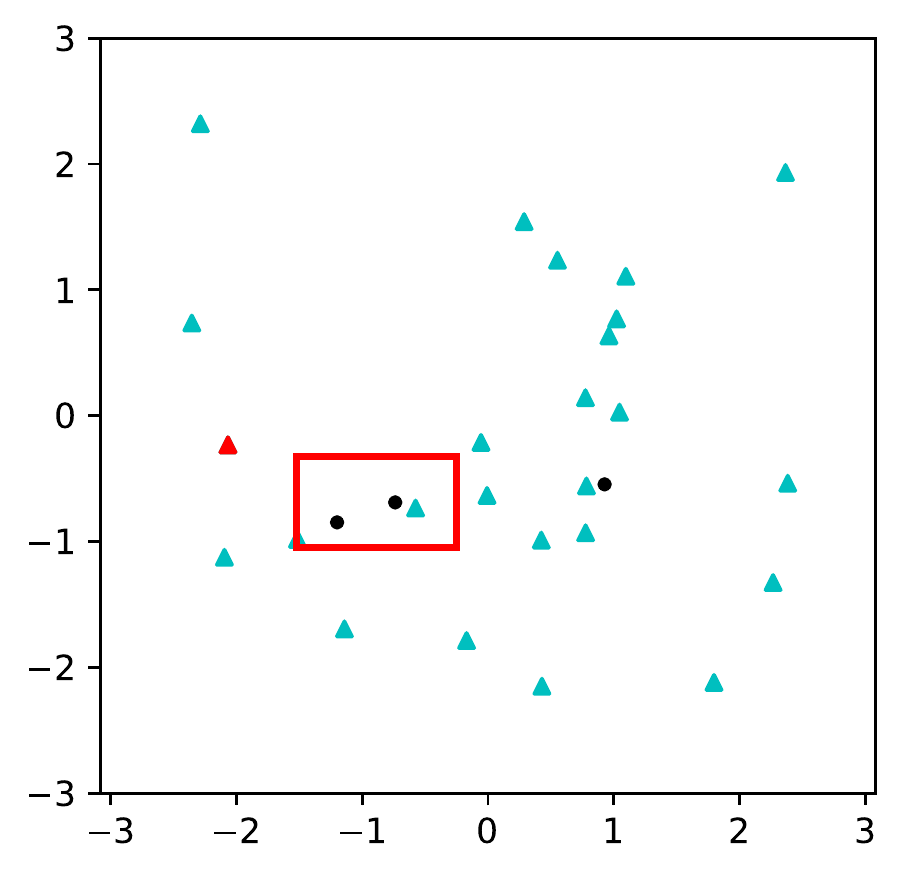}\\
		\end{tabular}
\vspace{-0.2cm}
\caption{Visualization of CubeRec-G (left) and CubeRec-A (right) for the same user group and recommended positive item on Meetup. Note that only the first two dimensions of $\mathbb{R}^d$ are used to facilitate visualization. The amount of visualized negative items is different as all representations are learned from two independent vector spaces, and embeddings out of the $[-3,3]$ range are clipped out to ensure clarity.}
\label{Figure:visualization}
\vspace{0.2cm}
\end{figure}

\subsection{Recommendation Effectiveness (RQ1)}
Table \ref{table:recommendation} summarizes the performance comparison among all the group recommenders, where each method's results are averaged over five runs on every dataset. With the recommendation results, we discuss our key findings below.

Apparently, CubeRec-G consistently outperforms all baselines by a significant margin. Compared with GroupIM which is the best baseline, the average improvements on Recall@10 and NDCG@10 brought by CubeRec-G are 14.4\% and 23.6\%, respectively. Unlike the straightforward mean pooling from user embeddings in NCF-AVG and FM-AVG, the selective preference aggregation schemes in AGREE, SIGR, CAGR and GroupIM generally achieve better performance when learning group-level representations for recommendation. However, these baselines are still subject to limited flexibility and resilience of the point embeddings used for representing groups. In contrast, our hypercube-based group representations can bypass the inherent limitations of point embeddings, thus achieving the best recommendation effectiveness.

Meanwhile, although CubeRec-A is not the best-performing variant of CubeRec, it still yields highly competitive results which are on par or stronger than the best baseline GroupIM. On one hand, its strong performance further demonstrates the benefit of replacing point embeddings with the more expressive hypercube representations. On the other hand, a possible reason that CubeRec-G is more advantageous than CubeRec-A is that, the group hypercubes learned via geometric bounding has a higher tolerance to the multi-faceted user preferences, and are overall more comprehensive representations of group-level interests. To qualitatively compare CubeRec-G and CubeRec-A, in Figure~\ref{Figure:visualization}, we visualize the learned hypercubes of a randomly picked group from Meetup. For clarity, we only use the first two dimensions of $\mathbb{R}^d$. The hypercubes and users in two plots belong to the same group in the dataset, and the same positive item has been successfully recommended in the top-10 list in both evaluation cases. As the visualization suggests, the hypercube learned by CubeRec-G is a tighter and more inclusive representation of all its members' preferences. Also, CubeRec-G appears to be more capable of pulling the positive item into the group hypercube than CubeRec-A, leading to better recommendation performance.

Another observation is the relatively inferior performance of all methods on Douban compared with other datasets. Despite the dense user-item interactions for learning individual users' preferences, the recommendation performance is largely impaired by the sparse group-item interactions and the largest item set. It is worth mentioning that methods with self-supervised learning (i.e., GroupIM and CubeRec) are still able to leverage augmented supervision signals to learn quality group representations, providing the best recommendation results even for this challenging dataset.

\begin{table}[t!]
\caption{Ablation test with different model architectures (Recall@10 is demonstrated). Numbers in bold face are the best results from each model, and ``$\downarrow$" marks a severe (over $5\%$) performance drop compared with the best results.}
\vspace{-0.3cm}
\centering
\setlength\tabcolsep{1.5pt}
  \begin{tabular}{c|c|c|c|c|c}
    \hline
     Method & Architecture & Yelp & Douban & Gowalla & Meetup\\
     \hline
    \multirow{4}{*}{CubeRec-G} & Default & \textbf{0.0455} & \textbf{0.0243} & \textbf{0.0361} & \textbf{0.0437} \\
    \cline{2-6}
    & Remove SR & 0.0440{\,\,} & 0.0222$\downarrow$ & 0.0353{\,\,} & 0.0414$\downarrow$ \\
    & Point Distance & 0.0204$\downarrow$ & 0.0098$\downarrow$ & 0.0156$\downarrow$ & 0.0181$\downarrow$ \\
    & Remove SSL & 0.0442{\,\,} & 0.0232{\,\,} & 0.0342$\downarrow$ & 0.0407$\downarrow$ \\
    \hline
    \hline
    \multirow{4}{*}{CubeRec-A} & Default & \textbf{0.0417} & \textbf{0.0211} & \textbf{0.0343} & \textbf{0.0418} \\
    \cline{2-6}
    & Remove SR & 0.0391$\downarrow$ & 0.0209{\,\,} & 0.0326{\,\,} & 0.0371$\downarrow$ \\
    & Point Distance & 0.0202$\downarrow$ & 0.0099$\downarrow$ & 0.0159$\downarrow$ & 0.0163$\downarrow$ \\
    & Remove SSL & 0.0382$\downarrow$ & 0.0195$\downarrow$ & 0.0342{\,\,} & 0.0399{\,\,} \\
    \hline   
    \end{tabular}
\label{table:ablation}
\vspace{0.1cm}
\end{table}

\subsection{Ablation Study (RQ2)}
To better understand the performance gain from the core components proposed in CubeRec, we perform ablation analysis on different degraded versions of CubeRec. Table \ref{table:ablation} summarizes the recommendation outcomes in terms of Recall@10. In what follows, we introduce all variants and discuss the effectiveness of corresponding model components.

\begin{figure*}[t!]
\centering
%\vspace{-0.3cm}
\begin{tabular}{cccc}
	\vspace{-0.2cm}\hspace{-0.15cm}\includegraphics[width=1.9in]{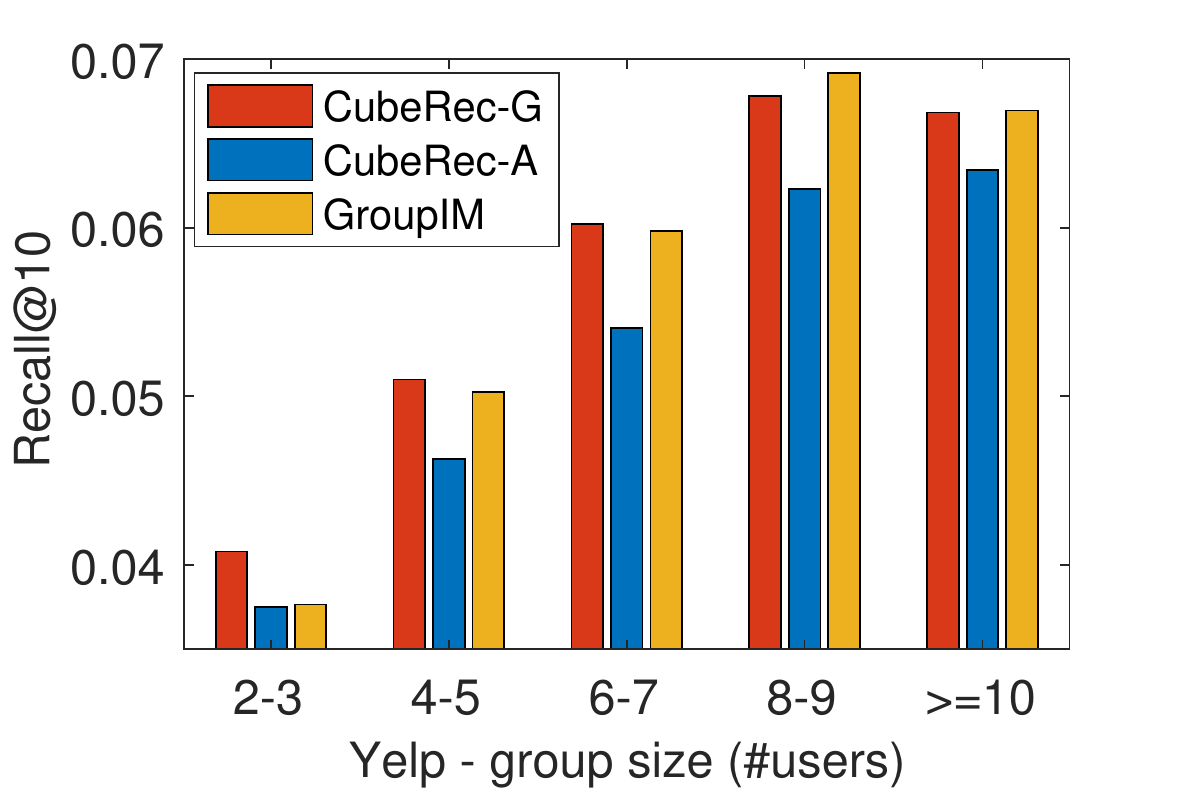}
	&\hspace{-0.7cm}\includegraphics[width=1.9in]{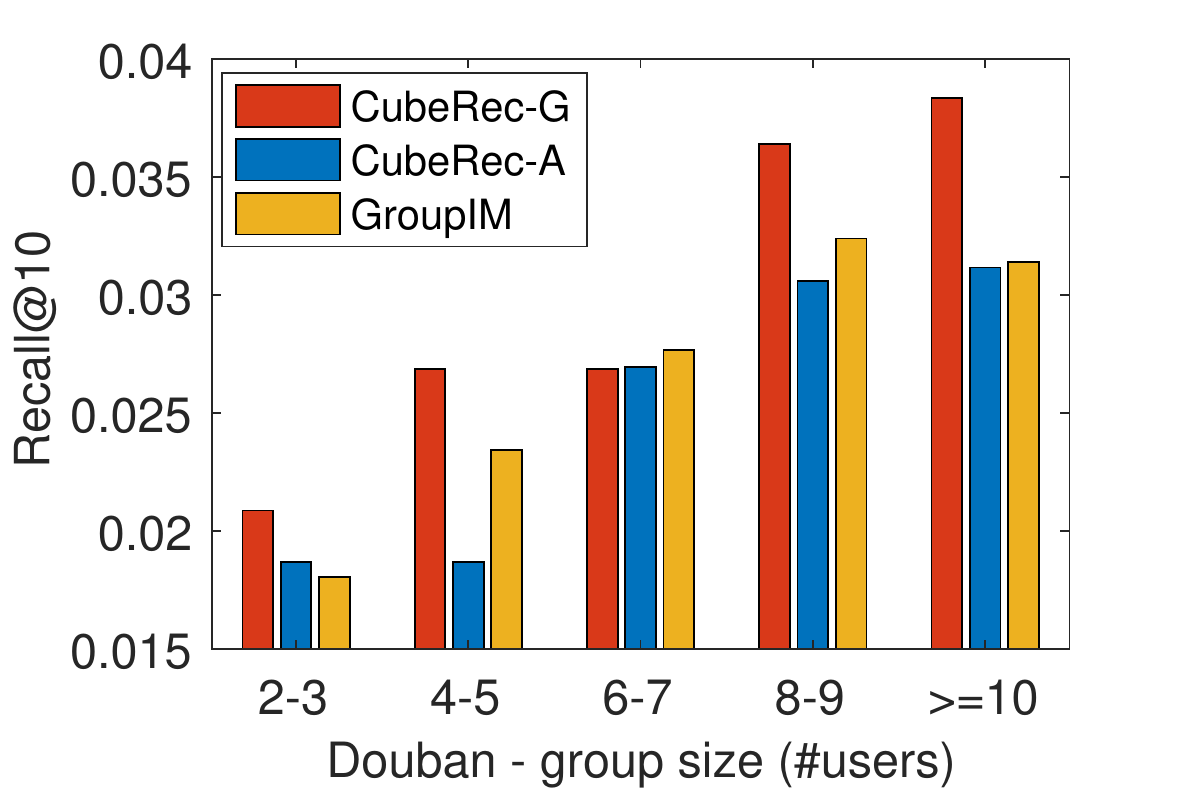}
	&\hspace{-0.7cm}\includegraphics[width=1.9in]{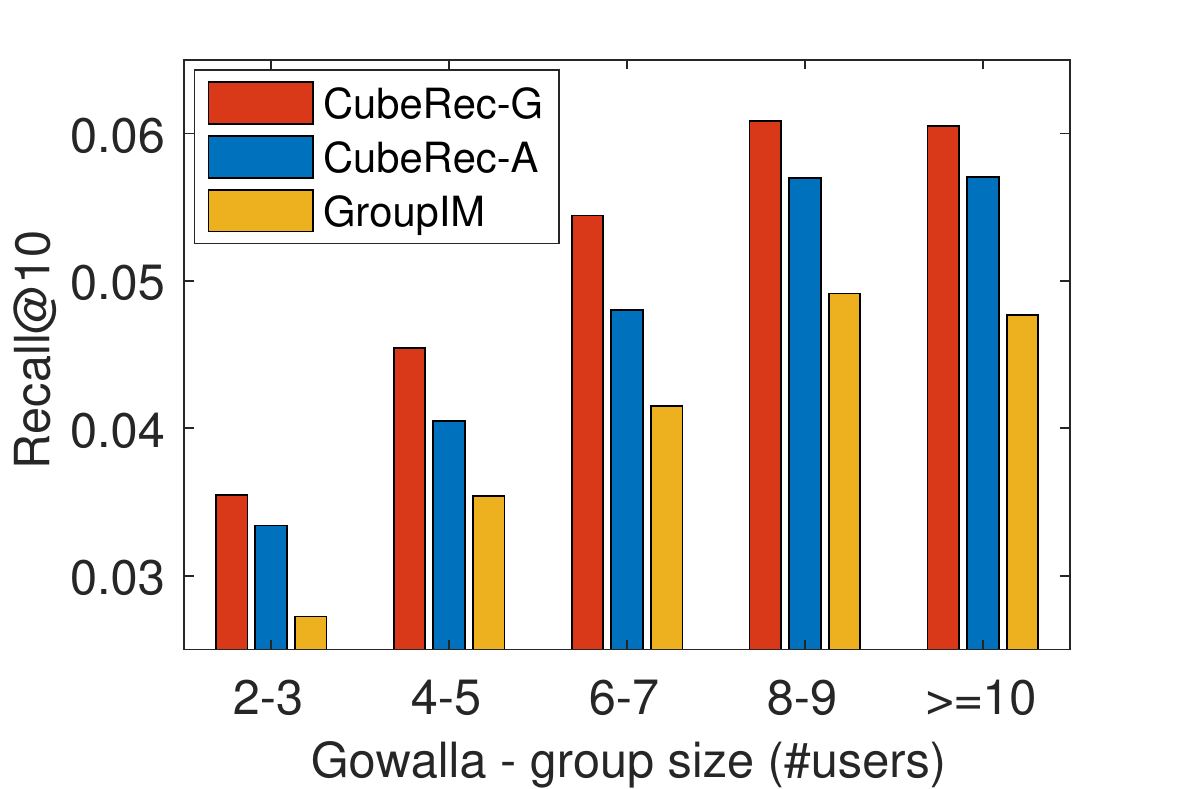}
	&\hspace{-0.8cm}\includegraphics[width=1.9in]{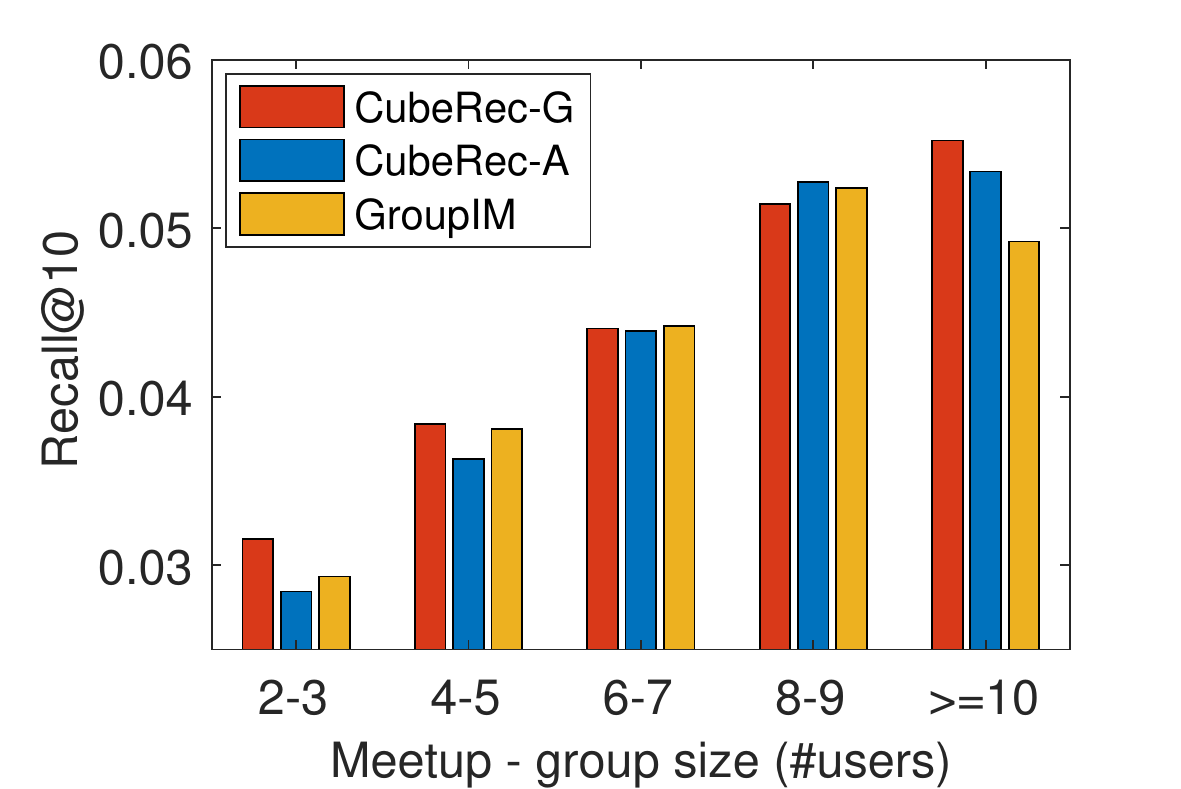}\\
		\end{tabular}
\vspace{-0.35cm}
\caption{Recommendation performance w.r.t. different group sizes.}
\label{Figure:group_size}
\vspace{-0.4cm}
\end{figure*}

\begin{figure}[t]
\centering
\renewcommand{\arraystretch}{0.1}
\begin{tabular}{ccc}
	\multicolumn{3}{c}{\includegraphics[width=2in]{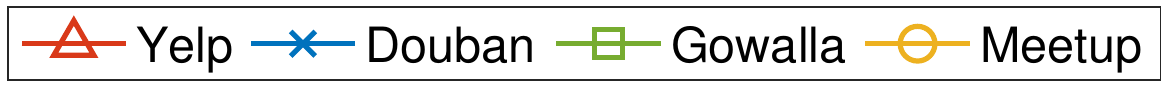}}\\
	\vspace{-0.2cm}\hspace{-0.2cm}\includegraphics[width=1.15in]{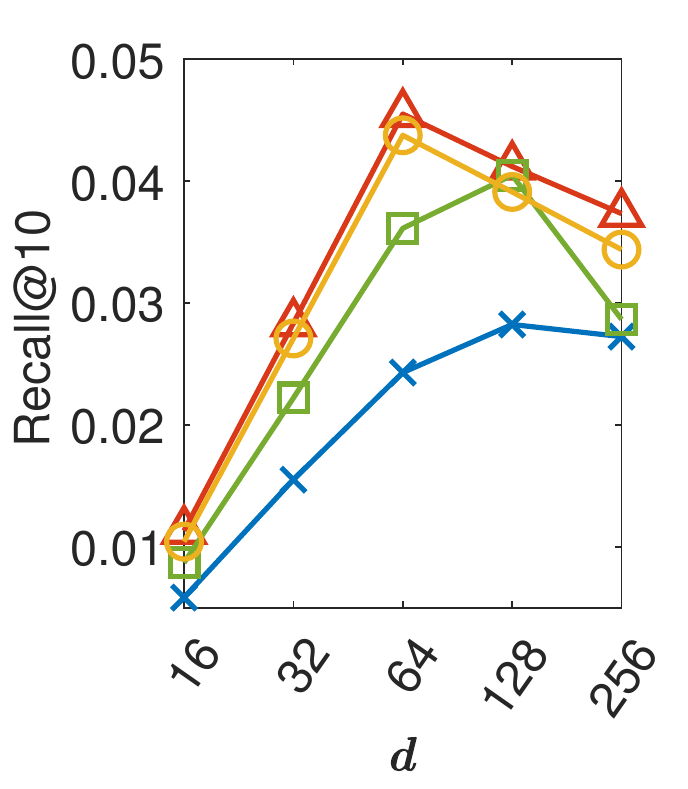}
	&\hspace{-0.4cm}\includegraphics[width=1.15in]{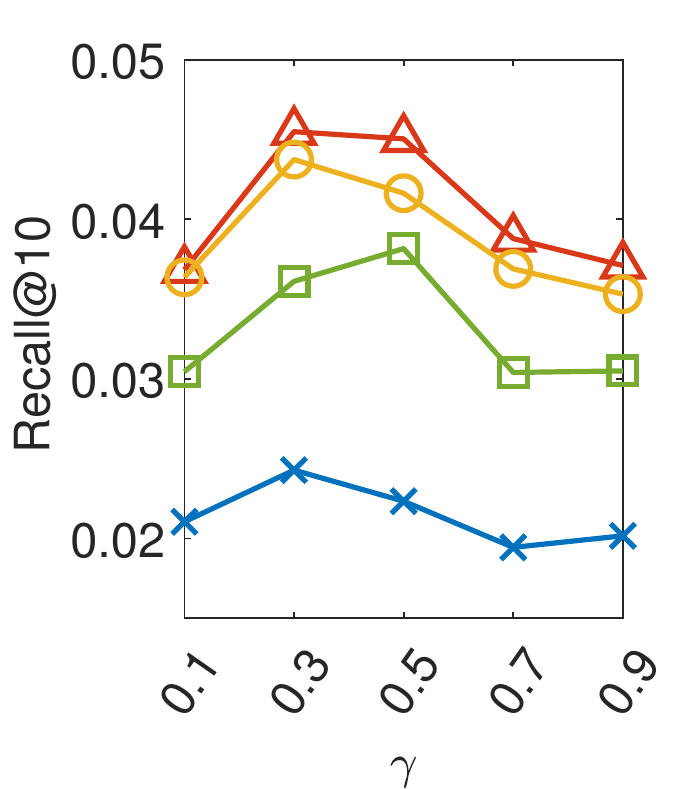}&\hspace{-0.45cm}\includegraphics[width=1.15in]{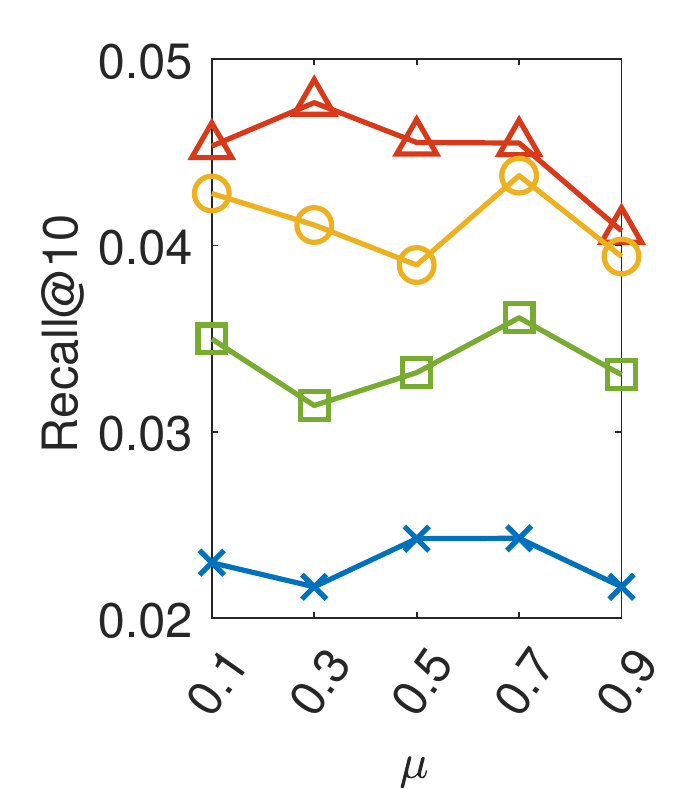}\\
		\end{tabular}
\vspace{-0.2cm}
\caption{Recommendation performance w.r.t. different hyperparameters.}
\label{Figure:paramsensitivity}
\end{figure}

\textbf{Removing Social Relations (Remove SR)}. In CubeRec, we infuse social relations when modelling user preferences via Eq.(\ref{eq:social_GCN}). To testify the usefulness of social relations in CubeRec, we remove the social relation matrix $\textbf{S}$ from the adjacency matrix by replacing it with a $|\mathcal{U}|\times|\mathcal{U}|$ matrix filled with 0s. The resulted variant models, have experienced noticeable performance drops for both CubeRec-G and CubeRec-A, especially on Meetup dataset with the highest sparsity of user-item interactions. Hence, the inclusion of social relations is important for CubeRec to learn representative user embeddings, which are the building block for group hypercubes.

\textbf{Using Only The Distance Between Item and Group Center Points (Point Distance)}. A crucial difference between CubeRec and point embedding-based group recommenders is the use of its hypercubes. Essentially, a group hypercube specifies a subspace in $\mathbb{R}^d$, where the recommended item points either locate in this subspace or maintain short distances with it. As a core ranking mechanism for group-item pairs in CubeRec, we verify the efficacy of such point-to-hypercube distance metric by replacing Eq.(\ref{eq:dnj}) with the conventional point distance. That is, $d_{nj}=||\textbf{c}_n-\textbf{v}_j||^2_2$ where squared Euclidean distance between the group center and item points are used. As a result, there is a significant decrease in recommendation for CubeRec across all datasets. This reflects that our revamped distance metric in CubeRec is a highly distinctive measurement for the affinity between item and groups. 

\textbf{Removing Self-supervised Learning (Remove SSL)}. This variant disables the self-supervised loss when optimizing CubeRec, i.e., the model is only optimized towards $L_{group}$ instead of $L_{group}+\mu L_{self}$. As CubeRec can no longer leverage the common users shared between groups as a supervision signal to counteract the group-level data sparsity and regularize the learned group hypercubes, it suffers from inferior recommendation accuracy on all datasets. Specifically, significant performance drops (over 5\%) are observed from CubeRec-G on Gowalla and Meetup, and from CubeRec-A on Yelp and Douban. As such, the novel self-supervised learning scheme showcases its strong contribution to the performance gain in CubeRec.

\subsection{Effect of Different Group Sizes (RQ3)}
In group recommendation, different group sizes will bring varied impact to the recommendation accuracy, and it is crucial for a recommender to keep robust in different circumstances. In this study, we further evaluate the performance of CubeRec on various group sizes. Following \cite{sankar2020groupim}, we segment all groups into 5 categories, namely groups with 2-3, 3-4, 6-7, 8-9, and 10 or more users. We additionally choose GroupIM as the peer method for comparison because it is the most performant baseline. 

We summarize the recommendation results in Figure \ref{Figure:group_size}, where Recall@10 is used for benchmarking. The first observation is that, CubeRec-G outperforms CubeRec-A and GroupIM in almost all scenarios. This also supports the superior performance of CubeRec-G given its capability to deal with different group sizes. Secondly, CubeRec-G yields more performance gain compared with GroupIM when the group sizes are relatively small (i.e., between 2 and 5). This is especially important for group recommendation, as the group sizes follow the typical long-tail distribution in all our datasets, considering it is more natural and feasible for users to form smaller groups in real life. Thirdly, we find that in general, group sizes are positively associated with the recommendation accuracy. Though smaller groups may have weaker discrepancies in user preferences during decision-making, the learning of group-level representations are also restricted by the limited user-item interactions. When the group size grows, the composed group representations tend to encode richer predictive signals for making accurate recommendations. Meanwhile, in some cases like Yelp and Gowalla, when the group size exceeds 9, the recommendation performance starts decreasing. This is largely associated with the excessive noise introduced by the diverse group members.  

\subsection{Hyperparameter Sensitivity (RQ4)}\label{sec:hyperparams}
We answer RQ4 by investigating the performance fluctuations of CubeRec with varying hyperparameters, in particular the latent dimension $d$, trade-off $\gamma$ between inner and outer distances in Eq.(\ref{eq:dnj}), and the self-supervision strength $\mu$. Based on the standard hyperparameter setup in Section \ref{sec:baselines_and_eval}, we tune the value of one hyperparameter while keeping the others unchanged, and record the new results achieved in Figure \ref{Figure:paramsensitivity}. Note that we only showcase the hyperparameter sensitivity for CubeRec-G since it is the best-performing of the two, and CubeRec-A exhibits a highly similar trend as CubeRec-G. 

\textbf{Impact of $d$}. The value of $d$ is examined in $\{16, 32, 64, 128, 256\}$. In general, CubeRec benefits from a relatively larger $d$ on all four datasets. But noticeably, the performance improvement stops when $d$ reaches a certain size (64 and 128 in our case) due to overfitting. 

\textbf{Impact of $\gamma$}. We also study the impact of $\gamma \in \{0.1, 0.3, 0.5, 0.7, 0.9\}$ that weighs the contribution of inner distance in our revamped group-to-item distance metric. As $\gamma$ increases from 0.1 to 0.3, there is a generally upward trend in CubeRec's performance. However, when $\gamma$ exceeds 0.5, the performance starts to decrease. Hence, in the revamped distance metric, accounting for slightly more outer distance offers higher resolution when distinguishing a group's preferences towards different items.  

\textbf{Impact of $\mu$}. The self-supervised learning loss is multiplied by $\mu$ to control its regularization effect. When the value of $\mu$ varies in $\{0.1, 0.3, 0.5, 0.7, 0.9\}$, the best performance of CubeRec is observed when $\mu=0.7$ in most scenarios, suggesting the necessity of utilizing our proposed self-supervision to counteract the sparsity of group-item interactions. However, CubeRec achieves better performance when $\mu=0.3$ on Yelp, which possibly attributes to the smallest average group size in this dataset (3.11 users per group), hindering the construction of self-supervision signals via relay users. 

\section{Related Work}\label{sec:related}
In general, group recommendation methods are designed in different ways to suit two types of audiences \cite{yin2019social,sankar2020groupim,quintarelli2016recommending}: persistent groups and occasional groups. While persistent group recommendation assumes dense interaction records at the group level, our work addresses occasional group recommendation, which is more practical and widely studied as real user groups tend to be formed ad-hoc with limited group-item interactions \cite{yin2019social}. Compared with persistent group recommendation where traditional collaborative filtering can be directly applied on group-item interactions \cite{hu2014deep,seko2011group}, a common practice in occasional group recommendation is to first learn individuals' preferences from user-level interactions, then perform preference aggregation to infer the overall interest of a group. Besides, a widely acknowledged challenge \cite{sankar2020groupim,yin2020overcoming} in occasional group recommendation is the highly sparse group-level interactions. Hence, we review two lines of research that contributes to group recommendation, namely methods for preference aggregation and counteracting data sparsity. 

\subsection{Preference Aggregation}
In group recommendation, all group members' preferences can be aggregated via either \textit{late aggregation} or \textit{early aggregation}. The aim of late aggregation \cite{amer2009group} is to generate item recommendations or predicted item-wise affinity scores for each group member, which are combined at the output stage via predefined strategies to produce recommendation results for the group. In this category, the most representative approaches are average satisfaction, least misery and maximum pleasure \cite{amer2009group,baltrunas2010group,quintarelli2016recommending}. 
 Unfortunately, these predefined aggregation strategies heavily rely on heuristics, thus lacking the desired expressiveness and flexibility for optimal performance \cite{guo2020group,jia2021hypergraph}.

As such, recent group recommenders mostly rely on early aggregation methods \cite{cao2018attentive,liu2012exploring,vinh2019interact,yin2020overcoming,yuan2014generative} for learning group-level preference representations. In contrast to late aggregation, such methods first aggregate
the preferences of all group members into a group-level representation, and then make group recommendations accordingly. Some early aggregation methods are built upon probabilistic models \cite{gorla2013probabilistic,liu2012exploring,yuan2014generative} by considering both a user's own preferences and her/his impact to the group, but the same type of probability distribution is assumed across users, which is infeasible in real-world cases. To address this problem, latent factor models \cite{sajjadi2021deepgroup,huang2020efficient} are adopted to map groups to embeddings in the vector space. For example, attentive neural networks are proposed in \cite{cao2018attentive,guo2020group} to selectively aggregate user representations within a group, and \cite{vinh2019interact} further captures the fine-grained interactions between group members via a sub-attention network. More recently, there are also studies to incorporate additional information like social connections \cite{cao2021social} and knowledge graphs \cite{deng2021knowledge} into the learning process of group representations. However, as discussed in Section \ref{sec:intro}, the point embeddings used in those methods sacrifice the diversity of personal preferences. Besides, compared with alternative solutions like capsule networks \cite{li2019multi,cen2020controllable} that rank items by mapping the learned multi-interest representations to a single embedding, CubeRec recommends items directly with group hypercubes, thus retaining loyalty to the decision-making process in real groups.

\subsection{Alleviating Data Sparsity for Groups}
When recommending items to occasional groups, the other crucial goal is to mitigate the scarce interactions between groups and items which can heavily hinder the quality of learned group representations. Some research account for the social relationships among users to provide hints on group-level preferences \cite{yin2019social,cao2021social,guo2020group}, such as \cite{yin2019social} that considers the impact of users' social influences on a group's decision. Afterwards, \cite{guo2020group} proposes an intra-group voting mechanism based on self-attention \cite{vaswani2017attention} to simultaneously model the social influence and the pairwise user interactions, and \cite{he2020game} further learns multi-view embeddings from the interactions among groups, users and items with a heterogeneous graph. As self-supervised learning (SSL) has shown its effectiveness in general recommendation tasks \cite{wu2021self,zhou2021pssl,yu2022self}, attempts are also made to design SSL objectives to counteract the data sparsity in group recommendation. For example, \cite{zhang2021double} leverages common users among groups and designs a double-scale contrastive learning to enhance the learned user representations. Meanwhile, \cite{sankar2020groupim} proposes a user-group mutual information maximization scheme to jointly learn informative user and group representations. Though SSL is proven to benefit group recommendation, existing methods are only compatible to point user/group embeddings in the vector space. In contrast, our design of SSL in CubeRec takes advantage of the geometric properties of hypercubes, allowing for more expressive group representations to be learned despite data sparsity.

\section{Conclusion}\label{sec:conclusion}
To learn high-quality group representations that have the capacity to encode the multi-faceted user preferences and the flexibility to fully resemble a group's decision-making process, we move beyond the traditional point embeddings, and propose to learn hypercube group representations in this paper. The proposed solution CubeRec takes advantage of its more expressive hypercubes in the vector space, and comes with a new distance metric and an intersection-based self-supervision paradigm to respectively facilitate group-item pairwise ranking and mitigation of data sparsity. Experimental results have rigorously verified the efficacy of CubeRec, shedding light on its applications in a wider range of recommendation tasks.

\section*{Acknowledgement}
T. Chen is supported by UQ New Staff Research Start-up Grant (NS-2103). This work is partly sponsored by Australian Research Council under the streams of Future Fellowship (FT210100624), Discovery Project (DP190101985) and Discovery Early Career Researcher Award (DE200101465). Y. Wang and M. Wang are supported by National Natural Science Foundation of China (U21A20470).

%\clearpage
%\newpage

\bibliographystyle{ACM-Reference-Format}
%\bibliography{bibliography}


\end{document}